\documentclass[prl,twocolumn,showpacs]{revtex4-1}
\usepackage{graphics}
\usepackage{graphicx}

\usepackage{amsmath}

\newcommand{\beq}{\begin{equation}}
\newcommand{\enq}{\end{equation}}
\newcommand{\beqa}{\begin{eqnarray}}
\newcommand{\enqa}{\end{eqnarray}}
\newcommand{\beit}{\begin{itemize}}
\newcommand{\enit}{\end{itemize}}
\newcommand{\bem}{\begin{pmatrix}}
\newcommand{\enm}{\end{pmatrix}}
\newcommand{\lat}{\left\langle}
\newcommand{\rat}{\right\rangle}
\newcommand{\av}[1]{\lat #1 \rat}

\newcommand{\eff}{\textrm{eff}}
\newcommand{\lb}{\left [}
\newcommand{\rb}{\right ]}
\newcommand{\lp}{\left (}
\newcommand{\rp}{\right )}
\newcommand{\Fab}{F_{\alpha\beta}}
\renewcommand{\bem}{\begin{bmatrix}}
\renewcommand{\enm}{\end{bmatrix}}
\begin{document}
\def\jnl@style{\it}
%commente par Seb
\def\aaref@jnl#1{{\jnl@style#1}}
%ref remplace par aaref pour eviter conflit...

\def\aaref@jnl#1{{\jnl@style#1}}

\def\aj{\aaref@jnl{AJ}}                   % Astronomical Journal
\def\araa{\aaref@jnl{ARA\&A}}             % Annual Review of Astron and Astrophys
\def\apj{\aaref@jnl{ApJ}}                 % Astrophysical Journal
\def\apjl{\aaref@jnl{ApJ}}                % Astrophysical Journal, Letters
\def\apjs{\aaref@jnl{ApJS}}               % Astrophysical Journal, Supplement
\def\ao{\aaref@jnl{Appl.~Opt.}}           % Applied Optics
\def\apss{\aaref@jnl{Ap\&SS}}             % Astrophysics and Space Science
\def\aap{\aaref@jnl{A\&A}}                % Astronomy and Astrophysics
\def\aapr{\aaref@jnl{A\&A~Rev.}}          % Astronomy and Astrophysics Reviews
\def\aaps{\aaref@jnl{A\&AS}}              % Astronomy and Astrophysics, Supplement
\def\azh{\aaref@jnl{AZh}}                 % Astronomicheskii Zhurnal
\def\baas{\aaref@jnl{BAAS}}               % Bulletin of the AAS
\def\jrasc{\aaref@jnl{JRASC}}             % Journal of the RAS of Canada
\def\memras{\aaref@jnl{MmRAS}}            % Memoirs of the RAS
\def\mnras{\aaref@jnl{MNRAS}}             % Monthly Notices of the RAS
\def\pra{\aaref@jnl{Phys.~Rev.~A}}        % Physical Review A: General Physics
\def\prb{\aaref@jnl{Phys.~Rev.~B}}        % Physical Review B: Solid State
\def\prc{\aaref@jnl{Phys.~Rev.~C}}        % Physical Review C
\def\prd{\aaref@jnl{Phys.~Rev.~D}}        % Physical Review D
\def\pre{\aaref@jnl{Phys.~Rev.~E}}        % Physical Review E
\def\prl{\aaref@jnl{Phys.~Rev.~Lett.}}    % Physical Review Letters
\def\pasp{\aaref@jnl{PASP}}               % Publications of the ASP
\def\pasj{\aaref@jnl{PASJ}}               % Publications of the ASJ
\def\qjras{\aaref@jnl{QJRAS}}             % Quarterly Journal of the RAS
\def\skytel{\aaref@jnl{S\&T}}             % Sky and Telescope
\def\solphys{\aaref@jnl{Sol.~Phys.}}      % Solar Physics
\def\sovast{\aaref@jnl{Soviet~Ast.}}      % Soviet Astronomy
\def\ssr{\aaref@jnl{Space~Sci.~Rev.}}     % Space Science Reviews
\def\zap{\aaref@jnl{ZAp}}                 % Zeitschrift fuer Astrophysik
\def\nat{\aaref@jnl{Nature}}              % Nature
\def\iaucirc{\aaref@jnl{IAU~Circ.}}       % IAU Cirulars
\def\aplett{\aaref@jnl{Astrophys.~Lett.}} % Astrophysics Letters
\def\apspr{\aaref@jnl{Astrophys.~Space~Phys.~Res.}}
                % Astrophysics Space Physics Research
\def\bain{\aaref@jnl{Bull.~Astron.~Inst.~Netherlands}} 
                % Bulletin Astronomical Institute of the Netherlands
\def\fcp{\aaref@jnl{Fund.~Cosmic~Phys.}}  % Fundamental Cosmic Physics
\def\gca{\aaref@jnl{Geochim.~Cosmochim.~Acta}}   % Geochimica Cosmochimica Acta
\def\grl{\aaref@jnl{Geophys.~Res.~Lett.}} % Geophysics Research Letters
\def\jcp{\aaref@jnl{J.~Chem.~Phys.}}      % Journal of Chemical Physics
\def\jgr{\aaref@jnl{J.~Geophys.~Res.}}    % Journal of Geophysics Research
\def\jqsrt{\aaref@jnl{J.~Quant.~Spec.~Radiat.~Transf.}}
                % Journal of Quantitiative Spectroscopy and Radiative Transfer
\def\memsai{\aaref@jnl{Mem.~Soc.~Astron.~Italiana}}
                % Mem. Societa Astronomica Italiana
\def\nphysa{\aaref@jnl{Nucl.~Phys.~A}}   % Nuclear Physics A
\def\physrep{\aaref@jnl{Phys.~Rep.}}   % Physics Reports
\def\physscr{\aaref@jnl{Phys.~Scr}}   % Physica Scripta
\def\planss{\aaref@jnl{Planet.~Space~Sci.}}   % Planetary Space Science
\def\procspie{\aaref@jnl{Proc.~SPIE}}   % Proceedings of the SPIE

\let\astap=\aap
\let\apjlett=\apjl
\let\apjsupp=\apjs
\let\applopt=\ao

\title{Information escaping the correlation hierarchy of the convergence field in the study of cosmological parameters.}
\author{Julien Carron}
\email[]{jcarron@phys.ethz.ch}
\affiliation{Institute for Astronomy, ETHZ, CH-8093 Zurich, Switzerland.}

\date{\today}

\begin{abstract}
%Using recent fits to numerical simulations, we show that the entire hierarchy of moments ceases to provide a complete description of the convergence one-point probability density function already for values of the associated matter fluctuations variance as low as 0.1, still in the weak lensing regime. This suggests that the full correlation function hierarchy of the convergence field becomes quickly generically incomplete and a very poor cosmological probe on non linear scales. At the scale of unit variance, only 5\% of the Fisher information content of the one-point probability density function is still contained in its hierarchy of moments, making clear that information escaping the hierarchy is a far stronger effect than information propagating to higher order moments. It follows that the constraints on cosmological parameters achievable through extraction of the entire hierarchy become suboptimal by large amounts. 
%A simple logarithmic mapping makes the moment hierarchy well suited again for parameter extraction, putting 80\% of the total information back into the first two and 95 \%  in the first three members. 

Using fits to numerical simulations, we show that the entire hierarchy of moments quickly ceases to provide a complete description of the convergence one-point probability density function leaving the linear regime. 
This suggests that the full N-point correlation function hierarchy of the convergence field becomes quickly generically incomplete and a very poor cosmological probe on nonlinear scales. At the scale of unit variance, only 5\% of the Fisher information content of the one-point probability density function is still contained in its hierarchy of moments, making clear that information escaping the hierarchy is a far stronger effect than information propagating to higher order moments. It follows that the constraints on cosmological parameters achievable through extraction of the entire hierarchy become suboptimal by large amounts. A simple logarithmic mapping makes the moment hierarchy well suited again for parameter extraction.
\end{abstract}
\maketitle

%-------------------------------- INTRODUCTION :
\paragraph{Introduction}

N-point correlation functions, first introduced in cosmology by Peebles and collaborators to describe the large scale distribution of galaxies \cite{1980lssu.book.....P}, are now ubiquitous in this field. They are at the heart of many cosmological probes like the CMB, galaxy clustering, or notably weak lensing, which was recognized as one of the most promising probe of the dark components of the universe \cite{2001PhR...340..291B,2004PhRvD..70d3009H,2006astro.ph..9591A,2009ApJ...695..652B}, and which traces the cosmological convergence field.\newline
 On large scales, or in the linear regime, correlations are a particularly convenient approach to tackle the difficult problem of statistical inference on cosmological parameters. Indeed, primordial cosmological fluctuation fields are believed to obey Gaussian statistics, and the first two members of the hierarchy, the mean and the two-point correlation function, provide a complete description of such fields. However, much less is known about the pertinence of the correlation hierarchy in the non-linear regime, or on small scales, where in principle a lot of information is contained, if only due to the large number of modes available for the analysis. More elaborated statistical models  must be made in this regime. For instance, the statistics of the matter field and its weighted projection the convergence field were shown to be closer to lognormal, at least in low dimensional settings \cite{1991MNRAS.248....1C,2000MNRAS.314...92T,2002ApJ...571..638T,2006ApJ...645....1D}, though with sizeeable deviations still.\newline
 Two effects relevant for statistical inference can in principle play a role entering the non linear regime, departing from Gaussian initial conditions. First, information may propagate to higher order correlators. Second, the correlation function hierarchy may not provide a complete description of the field anymore %\cite{footnote2}
 , so that information escapes the hierarchy. Even though this second possibility was pointed out qualitatively in an astrophysical context already in \cite{1991MNRAS.248....1C}, it seems it was not given further attention in the literature.
\newline
In this \textit{Letter} we show, using accurate fits of the convergence one-point probability density function to numerical simulations \cite{2006ApJ...645....1D} that the second effect very quickly completely dominates the convergence field, and thus that the hierarchy is not well suited for inference on cosmological parameters anymore. 
\paragraph{Fisher information and orthogonal polynomials.}
The approach is based on decomposing the Fisher's matrix valued information measure in components unambiguously associated to the independent information content of the correlations of a given order. It was recently proposed in \cite{2011ApJ...738...86C}, building upon \cite{Jarrett84}. Exact results at all orders were obtained only for the moment hierarchy of a idealized, perfectly lognormal one dimensional variable, where analytical methods could be applied. In cosmology, the Fisher information matrix is widely used for many years now to estimate the accuracy with which cosmological parameters will be extracted from future experiments aimed at some observables \cite[e.g.]{Tegmark97a,2004PhRvD..70d3009H,2009ApJ...695..652B}, assuming Gaussian statistics.

For a general probability density function $p(x,\alpha,\beta)$, $\alpha$, $\beta, \cdots$ any model parameters, its definition is
\beq \label{FI}
F_{\alpha\beta} = \av{\frac{\partial \ln p}{\partial \alpha}\frac{\partial \ln p}{\partial \beta}}.
\enq
Its inverse can be seen through the Cram\'er-Rao bound \cite{Tegmark97a} to be the best covariance matrix of the relevant parameters achievable with the help of unbiased estimators.
The general procedure to decompose the Fisher information content into uncorrelated pieces, corresponding to an orthogonal system, was presented in a statistical journal in  \cite{Jarrett84}.  When the observables of interest are products of the variables, i.e. moments or more generally correlation functions, the orthogonal system are orthogonal polynomials. It is discussed in detail in an cosmological context in \cite{2011ApJ...738...86C}.
In particular, the variables for which the Fisher information content on $\alpha$ is entirely within the first $N$ pieces, such as the Gaussian variables for $N = 2$, are those for which the function $\partial_\alpha  \ln p$ entering
(\ref{FI}), called the score function, is a polynomial of order $N$ in $x$.
In the case of a single variable, %\cite{footnote4}
the uncorrelated contribution of order $N$ to the Fisher information matrix $\Fab$ is given by
 \beq
s_N(\alpha)s_N(\beta),
 \enq
 where the Fisher information coefficients $s_N$ are the components of the score function with respect to the orthonormal polynomial of order $N$,
 \beq \label{sndef}
 s_N(\alpha) = \av{\frac{\partial \ln p}{\partial \alpha}P_N(x)},
 \enq
 \beq
 \av{P_n(x)P_m(x)} = \delta_{mn},\quad n,m \ge 0
 \enq
 For any $N$, the following relation holds
 \beq \label{sumsn}
 \sum_{n = 1}^Ns_n(\alpha)s_n(\beta) = \sum_{i,j = 1}^{N}\frac{\partial m_i}{\partial \alpha } \lb \Sigma^{-1}\rb_{ij}\frac{\partial m_j}{\partial \beta },
 \enq
 where $m_i = \av{x^i}$ and $\Sigma_{ij} = m_{i+j} - m_im_j$ is the covariance matrix.
 The right hand side being the expression describing the Fisher information content of the moments $m_1$ to $m_N$. Whether one recovers the full matrix $\Fab$ with $N\rightarrow \infty$ or only parts of it depends on the distribution under consideration. A sufficient condition is that the polynomials $P_n$ form a complete basis set, which is then essentially equivalent to the condition that the distribution can uniquely be recovered from its moments hierarchy \cite[and references therein]{1991MNRAS.248....1C,2011ApJ...738...86C}. This and other sufficient criteria for completeness are tightly linked to the decay rate of the probability density function at infinity. 
 
\begin{figure}
  \includegraphics[width = 0.5\textwidth]{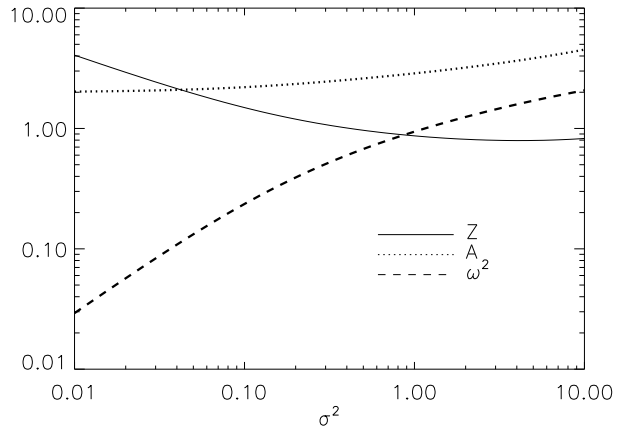}
 \caption{\label{fig1}The three parameters $Z$,$A$ and $\omega^2$ entering the generalized lognormal model, as function of the variance of $\delta_m^\eff $.} 
 \end{figure}

\begin{figure}
  \includegraphics[width = 0.5\textwidth]{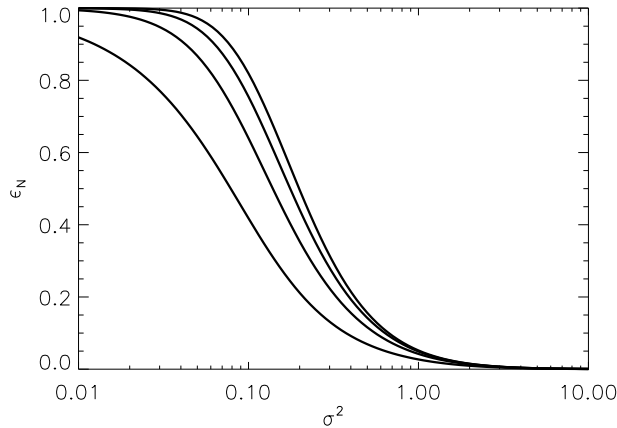}
 \caption{\label{fig2} The cumulative efficiency $\epsilon_N$ of the moments of the convergence in capturing Fisher information, for $N =2$ to $N = 5$, defined in eq. (\ref{eff}), from bottom to top, as function of the variance  of $\delta_m^\eff $.} 
 \end{figure}

\begin{figure}  \includegraphics[width = 0.5\textwidth]{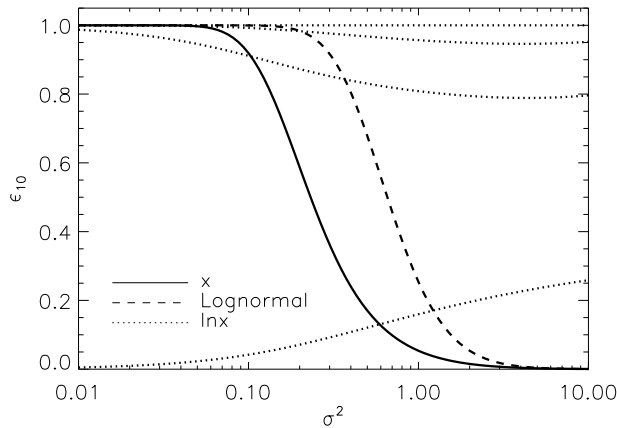}
 \caption{\label{fig3}Solid and dashed line : the efficiency $\epsilon_{N = 10}$ of the first $10$ moments to capture Fisher information, for the convergence field (solid) and lognormal field (dashed). The curves do not change anymore with increasing $N$. Dotted : $\epsilon_1$,$\epsilon_2$, $\epsilon_3$ and $\epsilon_{10}$ for the logarithmic transform of the field, from bottom to top.} 
 \end{figure}

\begin{figure}
  \includegraphics[width = 0.5\textwidth]{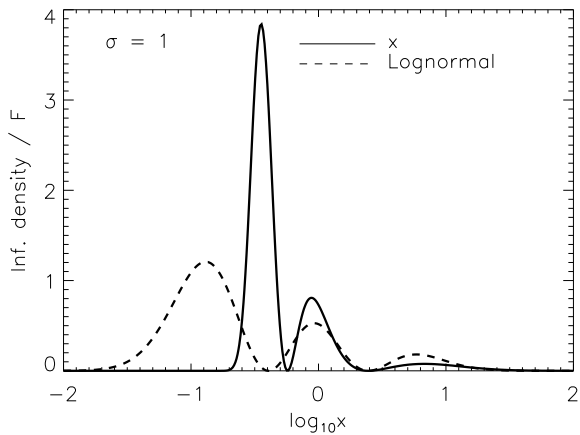}
 \caption{\label{fig4} The Fisher information density of the lognormal (dashed) and the convergence field (solid), renormalized such that it integrates to unity. Clearly, the Fisher information is mostly contained within the underdense regions. The moments are however sensitive to the tail.}
  \end{figure}

 We define the cumulative efficiency $\epsilon_N$ of the moments up to order $N$ to capture Fisher information on $\alpha$ as
 \beq
 \label{eff}
 \epsilon_N(\alpha):= \frac{\sum_{n = 1}^Ns_n^2(\alpha)}{F_ {\alpha\alpha}}.
 \enq
 From the Cram\'er-Rao bound, $\sqrt{\epsilon_N}$ is the ratio of the the best constraints achievable on $\alpha$ with any unbiased estimator to the expected constraints on $\alpha$ from the extraction of the first $N$ moments .
 \paragraph{Fisher information coefficients}
 We use the fits to simulations from \cite{2006ApJ...645....1D}, valid down to the arcsecond scales.  Initially built to correct for the failure of the lognormal distribution to reproduce the high and low density tails of the convergence $\kappa$ on a single lens plane, it reproduces accurately the cosmological convergence as well, taking into account the broader lensing kernel \cite{2011ApJ...742...15T}. In terms of the reduced variable $x$,
  \beq
 x = 1 + \frac \kappa{ \left |\kappa_{\textrm{empty}} \right |} =: 1 + \delta_m^\eff,
 \enq
 where $\kappa_\textrm{empty}$ is the minimal value of the convergence, corresponding to a light ray traveling an empty region, it takes the form of a generalized lognormal model for the associated effective matter fluctuations $\delta_m^\eff$,
 \beq \label{model}
 p(x,\sigma) = \frac{Z}{x} \exp \lb -\frac{1}{2\omega^2} \lp \ln x + \frac{\omega^2}{2}   \rp^2\lp 1 +\frac{A}{x} \rp \rb.
 \enq
 In this equation, the three parameters $Z$, $A$ and $\omega$ are such that the mean of $x$ is unity, and its variance $\sigma^2 = \sigma^2_\kappa / \kappa^2_{\textrm{empty}}$ (we are neglecting here a small but non-zero mean of the convergence argued in \cite{2011ApJ...742...15T}). Therefore, the only relevant parameter is the variance of the associated matter fluctuations $\sigma^2$, fixed by the cosmology from $\kappa_{\textrm{empty}}$ and the convergence power spectrum, together with some filter function corresponding to the smoothing scale, determining the level of non linearity of the field \cite[figure 1 ]{2006ApJ...645....1D}. Linear and non-linear regime being separated at $\sigma^2 \approx 1$. We obtained %\cite{footnote1}
 $Z,A$ and $\omega^2$, shown in figure \ref{fig1}, 
 with the help of a standard implementation of the Newton-Raphson method for non-linear systems of equations.
 %With the probability density function at hand, there are several ways to obtain the Fisher information coefficients $s_N(\sigma^2)$. One is through the evaluation of the moments and their derivatives and the use of equation (\ref{sumsn}). This is however hardly to be recommended due to the very severe ill-conditioning of moment and covariance matrices, even for moderate value of $N$.
 \newline
Orthogonal polynomials can very conveniently be generated by recursion, as exposed in details in \cite{Gautschi04}, since they satisfy a three terms recurrence formula. We define for convenience
 \beq
 \hat \pi_N(x) := \sqrt{p(x,\sigma)}\pi_N(x), 
 \enq
 where $\pi_N(x)$ is $P_N(x)$ rescaled such that the coefficient of $x^N$ is unity. The recursion relations in \cite {Gautschi04} %[theorem 1.27 ]
 become
  \beq
 \begin{split}
\hat \pi_{k + 1} &= (x - \alpha_k)\hat\pi_k -\beta_k \hat\pi_{k-1}, \\
 \alpha_k &:= \frac{\int_0^\infty dx\: x \:\hat \pi^2_k(x) }{\int_0^\infty dx\: \hat \pi^2_k(x) } \\
 \beta_k &:= \frac{\int_0^\infty dx\:\hat \pi^2_k(x) }{\int_0^\infty dx\:\hat \pi^2_{k-1}(x) },
 \end{split}
 \enq
 and $\pi_{-1}(x) = 0, \pi_1(x) = 1, \beta_0 = 1$, that we implemented using an appropriate discretization of the $x$-axis.  Proper normalization of the polynomials can be performed afterwards. The Fisher information coefficients were then obtained with the help of equation (\ref{sndef}), using a precise five point finite difference method for the derivatives of $Z,A$ and $\omega^2$ with respect to $\sigma^2$ that are needed to obtain the score function.\newline
In figure \ref{fig2}, we show the cumulative efficiency $\epsilon_N(\sigma^2)$, for $N = 2$ to $N = 5$, from bottom to top. (Note that $s_1(\sigma^2)$ vanishes since the mean of $x$ is unity for any value of the variance). The uppermost line contains therefore the variance, the skewness, the kurtosis as well as the 5th moment of the field.
The contribution of each successive moment can be read out from the difference between the corresponding successive curves. For higher $N$ quick convergence of $\epsilon_N$ occurs, presented in figure \ref{fig3} as the solid line, showing $\epsilon_{10}$. For small values of the variance, the field is still close to Gaussian, so that the Fisher information is close to be entirely within the the 2nd moment, and accordingly the ratio $\epsilon$ is close to unity in this regime. It is obvious from these figures that the main effect for larger values of the variance is not that Fisher information is transferred to higher order moments, but rather the dramatic cutoff as soon as the variance crosses $0.1$.  At redshift 1, this corresponds to the scale of $\approx 1'$ \cite[figure 1 ]{2006ApJ...645....1D}, so still within scales probed by weak lensing. For $\sigma \sim 1$, the ratio is close to $0.05$, meaning that all moments completely fails to capture the information. Optimal constraints on any cosmological parameter entering $\sigma$ are thus for this value of the variance a factor $1/\sqrt{0.05} \sim 4.5$ tighter than those achievable with the help of the entire hierarchy. 
 \newline\noindent
In figure \ref{fig3} we compare these results to the exact analytical expressions given in \cite{2011ApJ...738...86C} for the lognormal distribution, shown as the dashed line. These are given by, accounting for the different normalization,
\beq
s^2_N(\sigma^2) = q^2 \frac{q^N}{1-q^N} \lp \prod_{n = 1}^{N-1} \lp1 - q^n \rp \rp\lp \sum_{n = 1}^{N-1} \frac{q^n}{1-q^n}\rp^2,
\enq
with $q := 1/ (1 + \sigma^2)$. The total Fisher information content being in this case $(q/\ln q)^2 / 2 -q^2/ (4\ln q)$.
There also the information content of the moments saturates quickly as $N$ grows. It is striking that the incompleteness of the moment hierarchy occurs much earlier in the convergence field than in the lognormal. This can be understood from the following considerations. The main effect of the improved model (\ref{model}) for the convergence is to reproduce accurately the very sharp cutoff of the probability density function at low convergence values \cite[figure 3-6]{2006ApJ...645....1D}. This cutoff is very sensitive to the variance of the field, more sensitive than the cutoff of the lognormal. However, there the contribution to the moment $m_n$, $x^n$, is beaten down by orders of magnitude. To make this point clearer, we show in figure \ref{fig4} the Fisher information density $p \lp \partial_{\sigma^2}\ln p\rp^2$ for the lognormal distribution (dashed) and the model we used (solid), at the scale of $\sigma = 1$ It is obvious in both cases that a large fraction of the information is contained in the underdense regions, describing the cutoff of the distribution, but unaccessible to the moments of $x$. Since this is even more the case for the convergence field, the efficiency is accordingly even worse.
\paragraph{Restoration of the information} Finally we investigate to what extent the moment hierarchy of $\ln x$ contains more Fisher information than the hierarchy of $x$. Though our method is completely independent, this can be seen as complementary to recent works looking at the statistics of the field after local transforms, and at the statistical power of its power spectrum initiated in \cite{2009ApJ...698L..90N,2011ApJ...731..116N,2011ApJ...729L..11S}, even though in these works the fact that information actually completely escapes the hierarchy is not appreciated. This is done with the very same method used above, by obtaining the polynomials orthogonal to the distribution of $\ln x$, or equivalently decomposing the score function of $x$ in polynomials in $\ln x$ rather than in $x$. This is seen to perform very well, as shown by the dotted lines in figure \ref{fig3}. From bottom to top are plotted $\epsilon_1$,$\epsilon_2$ and $\epsilon_3$. Also shown in the figure is $\epsilon_{10}$ but it is not to be distinguished from unity, meaning that completeness of the hierarchy is restored. We see that over the full range at least $80\%$ of the information is back in the two first moments, and $95\%$ in the first three.
%-------------------------------- CONCLUSIONS :
\paragraph{Conclusions}
We have studied the statistical power of the moment hierarchy of the convergence field, when leaving the linear regime. Notably, the hierarchy ceases to provide a complete description of the statistics of the convergence, letting an increasingly large fraction of the Fisher information actually escape the hierarchy, and thus making constraints on cosmological parameters achievable with measurements of the hierarchy suboptimal by increasingly large factors. While our results are exact only for the one point distribution (or equivalently the full correlation function hierarchy of the convergence field in the limit of vanishing correlations), the correlation function hierarchy will also show a similar behavior, though the amplitude of the loss in information and constraining power may vary from parameter to parameter in the details. This is because this defect, for any number of variables, is due to the very slow decay rate at infinity of the field distribution, which cannot be reproduced by the exponential of a polynomial in the relevant variables. Our findings are consistent with previous analytical results on the lognormal distribution \cite{2011ApJ...738...86C}, and numerical work from $N$-body simulations at the power spectrum level \cite{2009ApJ...698L..90N,2011ApJ...731..116N}. Making a tighter connection to such simulation results with the methods presented here is the subject of future work. Of course, the quest for the information in the non linear regime already has problems of its own, such as shot noise issues, or accurate modeling, that we did not consider here. Nonetheless, these results clearly shows that if the correlation function hierarchy is to play a substantial role in getting constraints out of the mildly or non-linear regime, then an approach similar to a Gaussianizing transform \cite{2011ApJ...731..116N,2011ApJ...729L..11S}, in this work a simple logarithmic mapping, can hardly be avoided though the details still needs to be figured out. It is reassuring that this approach seems to work well to first order, and that first steps have recently already been taken in that direction in perturbation theory \cite{2011ApJ...735...32W}, for the matter field. Our work also points toward low convergence regions as carrying large amounts of information, though the importance of noise issues needs to be clarified in this regime. Thus, many promising ways have still to be explored to make profit of mildly and non-linear scales.

\begin{acknowledgments}We would like to thank Adam Amara and Simon Lilly for inspiring discussions, and acknowledge the support of the Swiss Science National Foundation.
\end{acknowledgments}

%Merlin.mbs v4.21 2009-07-09.
%\bibliographystyle{plain}

%

\end{document}